\documentclass[prx,twocolumn,floats,superscriptaddress,longbibliography]{revtex4-1}

\usepackage[colorlinks=true,urlcolor=blue,citecolor=blue,linkcolor=blue]{hyperref}

\usepackage{amsmath, amsthm, amssymb}
\usepackage{graphicx}
\usepackage{color}
\usepackage{multirow}

\usepackage[sort&compress]{natbib}
\setcitestyle{numbers,square}
\usepackage{hyperref}
\hypersetup{
        colorlinks=true,
}

\DeclareMathOperator{\im}{Im}

\newcommand{\eqnref}[1]{(\ref{#1})}
\newcommand{\ket}[1]{|#1\rangle}
\newcommand{\bra}[1]{{\left\langle{#1}\right\vert}}

\newcommand{\eval}[1]{\langle #1 \rangle}

\newcommand{\ord}[1]{\mathcal{O}(#1)}

\begin{document}

\title{Hybrid quantum-classical approach to correlated materials}

\author{Bela Bauer}
\affiliation{Station Q, Microsoft Research, Santa Barbara, CA 93106-6105, USA}

\author{Dave Wecker}
\affiliation{Quantum Architectures and Computation Group, Microsoft Research, Redmond, WA 98052, USA}

\author{Andrew J. \surname{Millis}}
\affiliation{Department of Physics, Columbia University in the City of New York, New York, NY 10027}

\author{Matthew B. Hastings}
\affiliation{Station Q, Microsoft Research, Santa Barbara, CA 93106-6105, USA}
\affiliation{Quantum Architectures and Computation Group, Microsoft Research, Redmond, WA 98052, USA}

\author{Matthias Troyer}
\affiliation{Theoretische Physik and Station Q Zurich, ETH Zurich, 8093 Zurich, Switzerland}

\begin{abstract}
Recent improvements in control of quantum systems make it seem feasible to finally build a quantum computer within a decade.
While it has been shown that such a quantum computer can in principle solve certain small electronic structure problems and idealized model Hamiltonians,
the highly relevant problem of directly solving a complex correlated material appears to require a prohibitive amount of resources.
Here, we show that by using a hybrid quantum-classical algorithm that incorporates the power of a small quantum computer into a framework of classical embedding algorithms, the electronic structure of complex correlated materials can be efficiently tackled using a quantum computer.
In our approach, the quantum computer solves a small effective quantum impurity problem that is self-consistently determined via a feedback loop between the quantum and classical computation.
Use of a quantum computer enables much larger and more accurate simulations than with any known classical algorithm, and 
will allow many open questions in quantum materials to be resolved once a small quantum computer with around one hundred logical qubits becomes available.
\end{abstract}

\maketitle

\section{Introduction}

The current workhorse for materials simulation is density-functional theory (DFT)~\cite{hohenberg1964}. DFT circumvents the exponential scaling of resources required to directly solve the electronic quantum many-body Hamiltonian by mapping the problem of finding the total energy and particle density of a system to that of finding the energy and particle density of non-interacting electrons in a potential that is a functional only of the electron density, and requiring self-consistency between the density and potential. The non-interacting electron and self-consistency problems are manageable on classical computers. While the universal functional that gives the dependence of the potential on the electron density cannot be efficiently evaluated~\cite{schuch2009computational}, several approximations to it are known empirically to be good; for example the local density approximation (LDA)~\cite{kohn1965}. LDA and related approximations yield reliable results for many weakly correlated materials, such as band insulators, metals, semiconductors and classes of biomolecules. However, many effects, including the intriguing physics of strongly correlated transition metal materials near a Mott transition~\cite{Imada98}, the phenomenon of high-temperature superconductivity~\cite{Bednorz:1986gk}, the properties of heme and other molecular complexes involving transition metals ~\cite{Kovaleva08}, and the complex physics of the actindes~\cite{actinides} are beyond the scope of DFT.

To go beyond DFT, one can attempt to exactly solve the quantum many-body problem using methods such as full configuration interaction (FCI), however because the required computational effort on classical computers scales exponentially in the number of orbitals, the method is limited to relatively small systems. In contrast, it has been shown that quantum computers can in principle solve certain electronic structure problems~\cite{lloyd1996universal,AspuruGuzik2005} in polynomial time. Recent advances towards building a small quantum computer~\cite{barends2014,corcoles2015,mourik2012} have led to increasing interest in what a small quantum computer could realistically simulate, and it has been shown that the simulation of small molecules~\cite{Whitfield:2011bz,wecker2014,hastings2014,poulin2014,babbush2015} and simplified model Hamiltonians~\cite{wecker2015} is within reach. However, in part because of the multiplicity of relevant interaction terms, the scaling of the currently known algorithms is not benign enough to allow naive direct simulations of complex correlated materials, for which thousands of electrons would have to be considered.

To make progress, the problem must therefore be simplified. One approach is to approximate the material with idealized model Hamiltonians, such as the Hubbard model~\cite{hubbard1963}, which have a sufficiently small number of interaction terms that they can easily be studied on quantum computers~\cite{wecker2015}. While capturing qualitative phenomena, such simple models do not offer a quantitative description of real materials.

\begin{figure}[tp]
  \centering
  \includegraphics[width=\columnwidth]{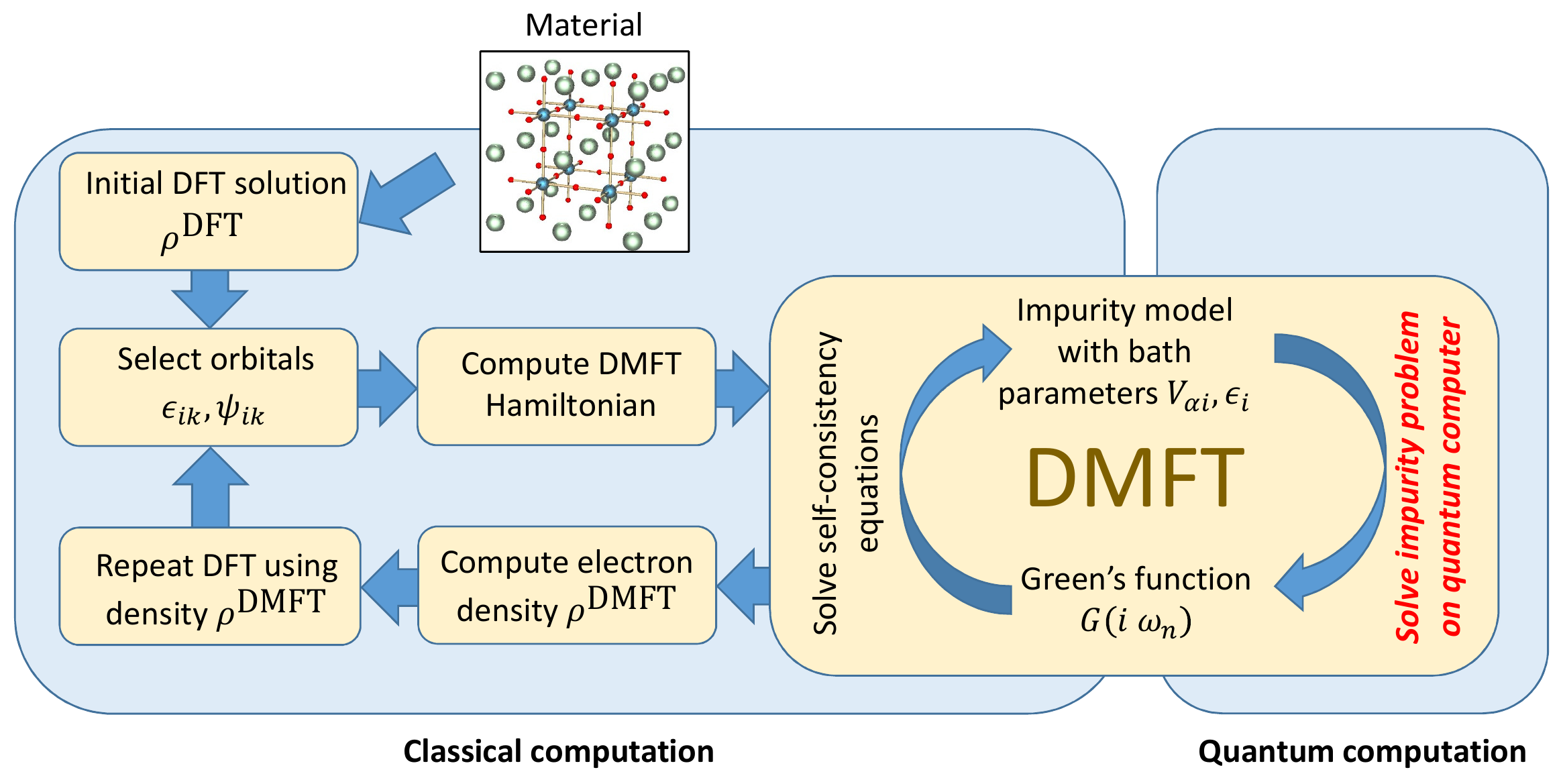}
  \caption{
  Overview of the DFT+DMFT approach. In our proposal, the solution of the impurity problem (highlighted in red), which is the computationally limiting step in computations using classical computers, is peformed by a quantum computer. \label{fig:overview} }
\end{figure}

\section{A hybrid quantum-classical approach}

We thus propose a hybrid approach, combining classical and quantum algorithms within the framework of the DFT plus dynamical mean field theory (DMFT) embedding approach. In this framework a computationally inexpensive DFT calculation is used to define a set of orbitals and determine the electronic structure of the majority of the orbitals, while a more expensive many-body method (here, DMFT) is used to solve a reduced model involving a much smaller set of correlated orbitals. The simplification lies in that one must deal only with a small set of correlated orbitals; the tradeoffs are that one requires a relatively complete solution of the correlated problem (full frequency dependence of the Green's function) and one obtains only an approximation to the true answer. The required highly accurate solution of a small problem is an ideal application for a small quantum computer where it enjoys maximal benefits over all known classical algorithms. 

This approach is being successfully employed in calculations of properties of correlated materials~\cite{kotliar2006}. DMFT provides an approximation to the solution of a full correlated problem by leveraging the solution of a {\em quantum impurity problem}, in which a finite cluster of interacting orbitals is self-consistently coupled to a bath of non-interacting electrons. DMFT becomes exact in an infinite lattice coordination limit or when the momentum dependence of the self energy may be neglected~\cite{metzner1989}, or when the number of orbitals in the impurity model becomes equal to the number of orbitals in the original problem to be solved \cite{Maier2005}. Practical implementations require impurity models with a relatively small number of orbitals, thus providing only an approximate solution to the full model; however in many cases the approximation is reasonably good. DMFT has been very successful at qualitatively describing the Mott transition~\cite{Georges:1992kt,georges1996} and its `cluster' extensions \cite{Maier2005} have produced important results for model systems (in particular the two dimensional Hubbard model). The combination with DFT is quantitatively explaining some properties of correlated materials~\cite{lda+dmft,kotliar2006}.

However, within a classical computational framework the complexity of the impurity model scales exponentially with the number of orbitals, placing severe limitations on the types of materials that can be tackled and restricting most real-materials DFT+DMFT simulations to the ``single-site'' DMFT approximation involving an impurity model representing a single correlated atom and neglecting all momentum dependence of the electron self energy. The restriction to just a small set of correlated orbitals also means that the method cannot be used to test the embedding hypothesis by systematically increasing the number of kinds of orbitals treated as ``correlated''. We show here that these limitations can be overcome by using a quantum computer to solve the impurity problem. Even an impurity problem with only $\sim 10^2$ degrees of freedom would enable the study of fundamentally new problems by allowing materials with multiple correlated atoms per unit cell to be considered, allowing cluster DMFT calculations of real materials and (given some further development of DMFT methodology) examination of the embedding hypothesis and the closely associated ``double counting correction''.

The Hamiltonian of the embedded impurity problem in DMFT may be written
\begin{align} \label{eqn:H} \begin{split}
H &= H_\mathrm{imp} + H_\mathrm{bath} + H_\mathrm{mix} \\
H_\mathrm{imp} &= \sum_{\alpha \beta} t_{\alpha \beta} c^\dagger_{\alpha} c_{\beta} +
	\sum_{\alpha \beta \gamma \delta} U_{\alpha \beta \gamma \delta} c^\dagger_\alpha c^\dagger_\beta c_\gamma c_\delta \\
H_\mathrm{mix} &= \sum_{\alpha i} \left(V_{\alpha i}  c^\dagger_\alpha d_i + \bar{V}_{\alpha i}  d^\dagger_i c_\alpha \right) \\
H_\mathrm{bath} &= \sum_i \epsilon_i d^\dagger_i d_i,
\end{split} \end{align}
where $c^\dagger_\alpha$ creates a fermion in one of the $N_\mathrm{so}$ spin orbitals of the interacting system labeled by a combined spin and orbital index $\alpha$; $d^\dagger_i$ creates a fermion on one of the $N_b$ bath sites. $N_{so}$ is finite by construction and $N_b=\infty$, but many approaches approximate the problem using a finite number of bath sites.

While the hopping integrals $t_{\alpha \beta}$ and interaction integrals $U_{\alpha \beta \gamma \delta}$ are directly given by the underlying material, the bath coupling $V_{\alpha i}$ and bath energies $\epsilon_i$ are determined from a self-consistency condition involving the Green's function of the impurity model and appropriate matrix elements of the Green's function of the lattice model. The self-consistency condition is typically solved iteratively by repeating the following steps:
(i) Starting with an initial guess for the bath parameters $\epsilon_i$ and $V_{\alpha i}$,
solve for the ground state of Eqn.~\eqnref{eqn:H} and extract the impurity Green's function $G$.
(ii) From $G$, calculate the self-energy $\Sigma = G_0^{-1} - G^{-1}$ and an updated non-interacting Green's function $G_0$.
(iii) Determine the discrete bath parameters $\epsilon_i$ and $V_{\alpha i}$ to closely match the desired non-interacting Green's function.
In practical calculations about twenty solutions of the impurity model are required. For further details, we refer to Appendix~\ref{app:selfc}.

The impurity solver is by far the most computationally demanding step in this loop. A variety of approaches exist~\cite{georges1996,gull2011}. Viewed as a generic quantum problem, the model has an interesting sparsity structure: the interactions only couple the $N_{so}$ impurity states, while the bath states are non-interacting. This sparsity structure is exploited by the widely-used continuous-time quantum Monte Carlo (QMC)~\cite{gull2011} methods, in which the bath is integrated out leaving an action involving $N_{so}$ degrees of freedom. However, these QMC algorithms~\cite{werner-millis,haule2007} scale exponentially in $N_\mathrm{so}$ and currently remain limited to $N_\mathrm{so} \approx 10$. Also they suffer from a severe sign problem in low symmetry situations where there is no choice of basis that diagonalizes the hybridization function
$\Delta_{\alpha\beta}(i \omega_n)=\sum_i V_{\alpha i} \bar{V}_{\beta i}/(i \omega_n - \epsilon_i)$ at all frequencies.

Exact diagonalization (ED) solvers \cite{krauth1994} approximate the continuous bath by a finite number of bath orbitals and do not take advantage of the sparsity structure. Therefore on a classical computer they have a cost that is exponential in the total system size $N_\mathrm{so}+N_d$. This and the need to obtain the full Green's function means that practical calculations are limited to $N_\mathrm{so}+N_d \approx 25$, in other words to five or fewer correlated orbitals, often corresponding to just a single correlated atom within a unit cell, and with a very small number of bath sites per correlated orbital. Recent developments \cite{Zgid12,Lu14,go2015} based on quantum chemical methods to define reduced basis sets for the ED calculation permit inclusion of somewhat larger numbers of bath orbitals, but at least as presently formulated these methods work in a natural orbital basis which strongly mixes the bath and correlated orbitals, so that the sparsity structure mentioned above cannot be exploited. In a parallel development, ideas to solve the impurity problem using tensor networks~\cite{garcia2004} have recently started to show great promise~\cite{wolf2014,ganahl2015,wolf2015}.

\section{Quantum algorithm for the impurity solver}

Significantly larger problems can be tackled when solving the impurity problem on a quantum computer. The key points are that the wavefunction of the impurity problem requires only $N_\mathrm{so}+N_d$ logical qubits and that the quantum computation takes advantage of the sparsity structure mentioned above. In particular, the number of bath sites affects the number of required qubits, but does not have a strong effect on the computation time. We leverage standard quantum algorithms discussed previously for quantum chemistry and model Hamiltonian applications to first obtain a quantum representation of the ground state of~\eqnref{eqn:H}, and then measure the Green's function.

To obtain the ground state, we combine adiabatic state preparation and quantum phase estimation (QPE)~\cite{kitaev1995,kitaev2002book}. We start from the easily prepared ground state of a simple Hamiltonian $H_0$ and evolve it under a time-varying Hamiltonian $H(t)$ that adiabatically interpolates from $H_0$ to the desired Hamiltonian~\eqnref{eqn:H}. Changing the parameters slowly compared to the inverse spectral gap of $H(t)$ ensures that the wave function always remains close to the ground state. Possible choices for the initial Hamiltonian could be either the atomic limit of turning off all hopping terms, such that the ground state becomes a simple product state of occupied and unoccupied spin orbitals, or turning off interactions such that the initial state is a Slater determinant which can be efficiently prepared using techniques discussed in Ref.~\onlinecite{wecker2015}. At the end of the adiabatic process, QPE can be used to measure the energy of the state and
collapse the wavefunction into an eigenstate $|\Psi\rangle$ of the Hamiltonian.
QPE relies on the ability to apply $\exp(-i H t)$ to the state, and avoids measuring the individual terms of the Hamiltonian
separately, since such measurements do not commute among themselves and with $H$ and would thus destroy the state. In contrast,
quantum phase estimation performs a measurement that is diagonal in the energy basis and which will project a state close to the ground state
onto the ground state with high probability. This is achieved with an accuracy $\epsilon \sim \mathcal{O}(1/T)$, where $T$ is the total computation time.
Details of the method are discussed in Appendix~\ref{app:algorithms}.
State preparation has succeeded if we measure the ground state energy ( $|\Psi\rangle$ then is the corresponding ground state wave function), but needs to be repeated if the measured energy corresponds to an excited state.

Due to the constraint of unitary evolution on a quantum computer, we can only measure the Green's function in real time (or real frequencies~\cite{wecker2015}). For $t \geq 0$, the particle and hole Green's functions in real time are defined as:
\begin{align} \begin{split} \label{eqn:rtgf}
G_{\alpha \beta}^p(t) =& \langle \Psi | c_\alpha(t) c_\beta^\dagger(0) | \Psi \rangle \\
G_{\alpha \beta}^h(t) =& \langle \Psi | c_\alpha^\dagger(t) c_\beta(0) | \Psi \rangle.
\end{split} \end{align}
In the following, we present the details involved in extracting the Green's function in Matsubara frequencies from our quantum impurity solver.
We will suppress the orbital indices for notational simplicity and implicitly assume that the Green's function is treated as a matrix.
As the self consistency condition is best enforced in imaginary frequencies, we perform a Hilbert transformation to find
\begin{equation} \label{eqn:Giw}
G(i \omega_n) = -i \int_{\epsilon}^\infty dt\ e^{-t \omega_n} \left[G^p(t) + \bar{G}^h(t) \right],
\end{equation}
where $\omega_n = \frac{\pi (2n+1)}{\beta}$, $n=1,...,N_\omega$ and $\beta$ is a fictitious inverse temperature. To obtain ground-state properties, $\beta$ is chosen sufficiently small to guarantee converged results and the frequencies are cut off at some suitably chosen $N_\omega$. In practice, the integral in \eqnref{eqn:Giw} is replaced by a discrete sum over a logarithmic grid, and the real-time Green's function must be measured separately for every time point. This is discussed in more detail in Appendix~\ref{app:selfc}. Alternative approaches to measure dynamical correlation functions are discussed in Ref.~\onlinecite{wecker2015}.

\begin{figure}
  \centering
  \includegraphics[width=\columnwidth]{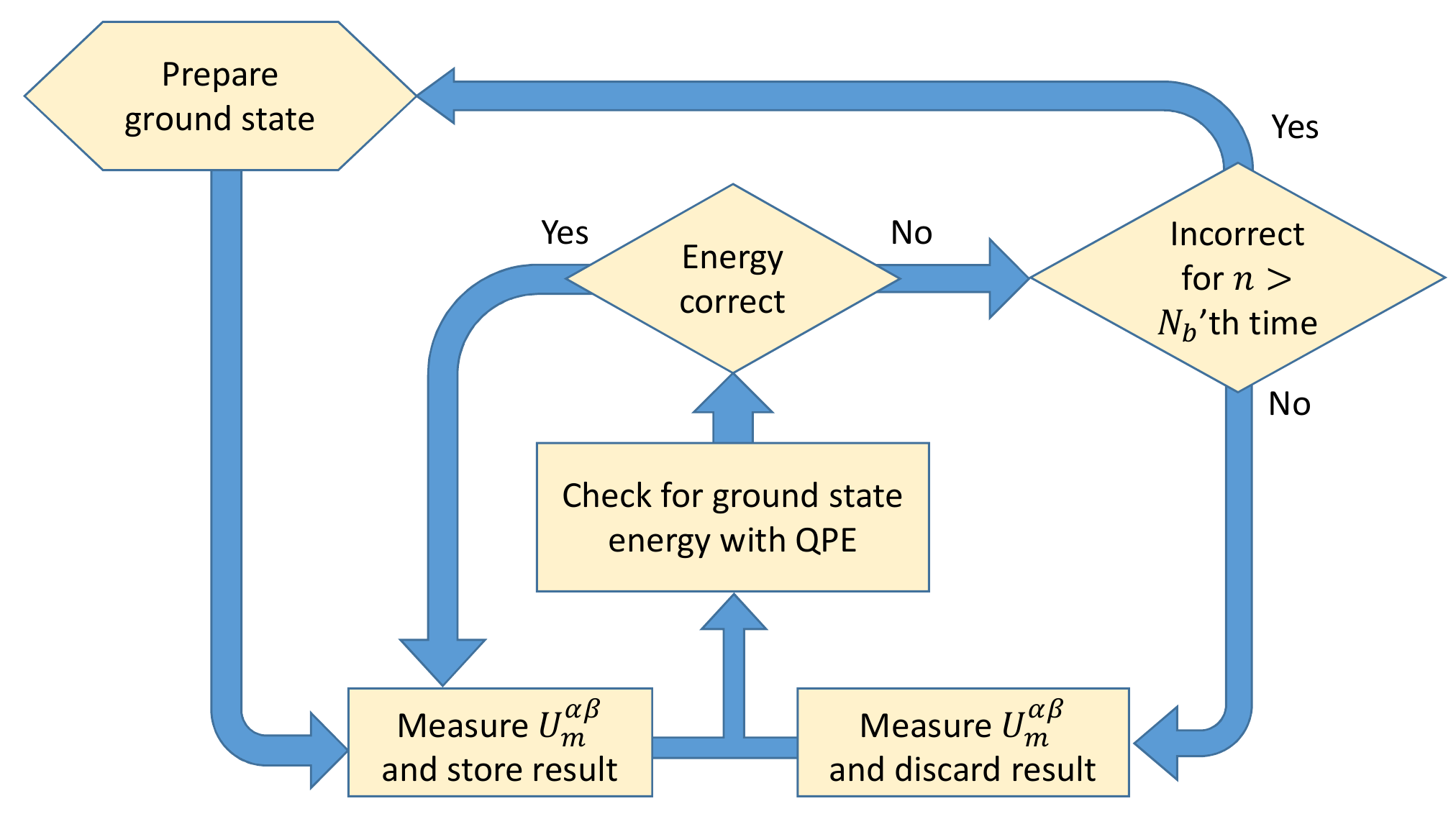}
  \caption{Overview of the incoherent estimation of the Green's function. \label{fig:incoherent} }
\end{figure}

To measure the real-time Green's functions~\eqnref{eqn:rtgf}, we relate them to expectation values of unitary
operators $q_1(t)=c(t)+c^\dagger(t)$, $q_2(t) = i (c(t)-c^\dagger(t))$, which can be measured directly; for details,
see Appendix~\ref{app:gfct}. Note that this formulation allows for straightforward determination of the superconducting components of the Green's function, at the cost of twice as many measurements. In a naive approach, measurements are destructive and $| \Psi \rangle$ must be re-prepared for each measurement.

\begin{figure}
  \includegraphics[width=\columnwidth]{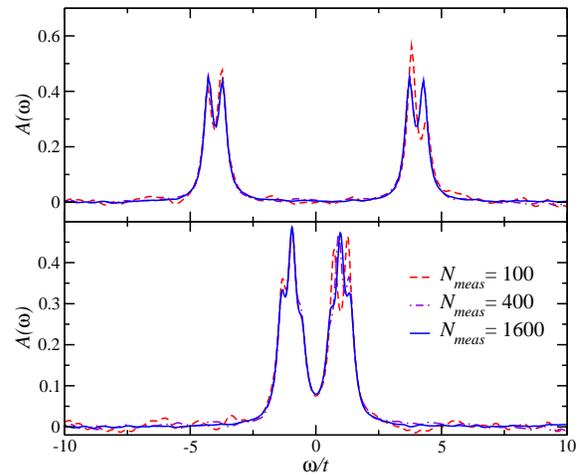}
  \caption{Spectral function of a Hubbard model on the Bethe lattice (with hopping $t_{ij}=t/\sqrt{2z}$ in the limit $z \rightarrow \infty$) with $U/t=8$ (upper panel)
  and $U/t=2$ (lower panel) and $N_b=10$ bath sites.
  For this simulation, both the imaginary-time Green's function required in the self-consistency as well as the spectral function were
  extracted from real-time Green's function data. For the self-consistency, a time grid of 1200 points from $T_\mathrm{min}=10^{-5}$ to
  $T_\mathrm{max}=40$ was used. Other parameters of the self-consistency were $\beta=20$, $N_\omega=400$. For extracting
  the spectral function, a time grid of 1000 points with the same limits was used. We emphasize that no analytic continuation is
  required to obtain $A(\omega)$ in our approach.
  $N_\mathrm{meas}$ in the legend indicates how many good samples of the real-time Green's function were obtained at
  each time slice. The results for $N_\mathrm{meas}=100$ are very similar to those for larger $N_\mathrm{meas}$ and those for $N_\mathrm{meas}=400$ are almost indistinguishable from those for $N_\mathrm{meas}=1600$.
  \label{fig:Aw} }
\end{figure}

We perform numerical simulations of our proposed quantum algorithms to establish a baseline of how many gates need to be executed to solve a simple impurity problem. Since these simulations on a classical computer scale exponentially in the size of the impurity system, we are limited to very small problems. We thus consider a single spinful impurity site ($N_\mathrm{so}=2$) coupled to a bath of 5 spinful sites ($N_b=10$). We run a self-consistent DMFT calculation, based on a simulation of our quantum algorithm, for a Hubbard model on the Bethe lattice for two different strength of the on-site interaction $U$ corresponding to the Fermi liquid and the Mott insulating regime, respectively. Our results for the spectral function of the converged DMFT solution are shown in Fig.~\ref{fig:Aw}.

To evaluate the integral~\eqnref{eqn:Giw} in each iteration of the DMFT loop, we measure the real-time Green's function on a grid of 1000 time points and take 400 measurements at each time point, where we need to measure both the particle and hole contributions, the imaginary and real part as well as spin components separately (unless the system is spin-degenerate). These numbers have been chosen such that the error from the measurement of the Green's function is small compared to the uncertainty in the bath fitting procedure, i.e. the limitations of DMFT with a small, discrete bath. With these choices, we need a total of $3.2 \cdot 10^6$ measurements, each giving one bit of information. For each measurement, we prepare the ground state, which we found to require $3 \cdot 10^5$ total gate operations in this instance.

For more complex problems, the preparation of the ground state will be much more costly, and should thus be avoided if possible. Since each measurement only extracts one bit of information, the state after the projective measurement may have significant overlap with the ground state. A relatively short quantum circuit, using a QPE, can then be used to project back into the ground state. This motivates the approach sketched in Fig.~\ref{fig:incoherent}. For the simple test case considered here, the cost of performing a QPE is comparable to the adiabatic state preparation, and thus no advantage can be gained by attempting to project back into the ground state after measuring one bit. In general, however, adiabatic state preparation will scale with the square of the inverse gap, $\mathcal{O}(1/\Delta^2)$, while the cost of QPE scales only roughly linearly, $\mathcal{O}(1/\Delta)$, leading to quadratic advantage of QPE over re-preparing the state. For large simulations, avoiding the preparation at the cost of performing QPE more often is therefore highly advantageous.

Further improvements can be gained with a fully coherent measurement procedure as described in Ref.~\onlinecite{wecker2015} and also reviewed in Appendix~\ref{app:gatec}. This procedure quadratically speeds up sampling, reducing the scaling of the time required for a given accuracy from $\mathcal{O}(1/\epsilon^2)$ to $\mathcal{O}((1/\epsilon)\log(1/\epsilon))$. In our numerical example given above, the coherent approach will reduce the number of measurements needed to achieve the same accuracy from 400 to $\sqrt{400}=20$ and thus yield roughly a ten-fold improvement. In other applications, where a higher accuracy is required, the improvement will be more significant.

Having established a baseline for the number of gates that must be executed, we now address the question of how
this number scales with the size of the impurity problem. The important contributions to the scaling are
(i) the number of terms in the Hamiltonian, which determines the number of gates required to perform a single
 time step of the evolution,
(ii) the number of measurements that must be taken,
(iii) the time that is required to accurately prepare the ground state,
and (iv) the time step required to reach the desired accuracy.

The number of terms in the Hamiltonian scales like $\ord{N_\mathrm{so}^4+N_b+N_b N_\mathrm{so}}$ (note that $N_b$ enters linearly while $\ord{N_\mathrm{so}}$ enters quartically; this is an example of exploiting the sparsity structure mentioned above). If gates can be executed in parallel, an even more favorable scaling is obtained by mapping the bath onto a set of $N_\mathrm{so}$ chains (rather than the ``star" topology used in~\eqnref{eqn:H}); this can be achieved by block-tridiagonalizing the quadratic bath terms using a Krylov approach~\cite{si1994}. Using this mapping, many terms can be executed in parallel, such that the scaling becomes independent of the size of the bath and scales as $N_\mathrm{so}^3$. The number of non-commuting terms in the Hamiltonian also modestly affects the required time step~\cite{poulin2014}.
The self-consistency condition requires measurements of the $N_\mathrm{so} \times N_\mathrm{so}$ Green's function matrix. However, in many cases orbital and spin symmetries of the system can be used to block-diagonalize the Green's function and thus reduce the number of independent measurements.

Since our algorithm spends significant time preparing the ground state, the requirements of the adiabatic state preparation play a crucial role in scaling. This generally depends on the minimal spectral gap along the chosen adiabatic path, which in turn will depend on the physical properties of the system in question. For gapped systems such as Mott insulators or superconductors where the order parameter fluctuations are gapped, the required preparation time is not expected to scale significantly with the size of the bath or the complexity of the impurity. Systems with very small gaps, such as Kondo systems, or in gapless regimes, such as Fermi liquids, likely pose greater challenges in particular for the accurate preparation of the ground state, where special care must be taken to find an optimal adiabatic path and choose a sufficiently long preparation time. However, it is important to note even for physical systems which are gapless, the finite size of the discrete impurity bath induces a non-zero gap in the problem we have to solve.

Taking the above scaling considerations into account, a relevant physical problem of 10 orbitals ($N_\mathrm{so}=20$) with the corresponding number of 60-100 bath sites seems within reach for a small quantum computer of about one hundred qubits. Such a problem would require on the order of $10^8$ measurements, which each can be achieved in a coherent run of about $10^8$ gates. While this leads to a large total number of gates of $10^{16}$, it is important to keep in mind that in contrast to other approaches~\cite{Whitfield:2011bz,wecker2014}, these gates need not be executed in a single coherent simulation, but are broken up into $10^8$ independent runs that can be executed sequentially or in parallel if several quantum computers are available. If the concrete quantum computing architecture allows the execution of more gates coherently, large improvements on the total gate count can be gained from exploiting the improved measurement schemes mentioned above.

\section{Outlook and Discussion}

Our hybrid quantum-classical approach to materials simulations is similar in spirit to complete active space methods in quantum chemistry, which pick a subset of orbitals to be treated by a method with higher accuracy, but go beyond these methods by feeding back the solution of the quantum impurity problem into the DFT problem. The ideas put forward here are not restricted to the commonly used DFT+DMFT approach, but can be generalized to other quantum embedding approaches such as the recently proposed density-matrix embedding theory (DMET)~\cite{knizia2012}. There, one strives to attain self-consistency between an extended non-interacting lattice model and an interacting impurity problem.
The parameters that must be determined self-consistently are hopping parameters of the non-interacting model, which are chosen in such a way that the single-particle equal-time Green's functions of the two models match.
This scheme requires knowledge only of equal time Green's functions which are straightforward to measure on a quantum computer.

While it has been known for a long time~\cite{Feynman1982,lloyd1996universal} that many quantum problems can be simulated on quantum computers with polynomial scaling, such scaling in the asymptotic regime is insufficient to make an algorithm practical, especially if the power of the polynomial is high and constants are large -- as is the case for quantum chemistry solutions of molecules and materials~\cite{Whitfield:2011bz,wecker2014}. Recent improvements in algorithms and runtime estimates~\cite{hastings2014,poulin2014,babbush2015} make the solution of small but classically challenging molecules practical on small quantum computers.

With our hybrid quantum-classical algorithm small quantum computers will also be useful for the simulation of larger systems, and especially strongly correlated crystalline materials or complex molecules which exhibit a wide variety of interesting physical phenomena and have a broad range of applications. Materials simulations are today one of the major uses of supercomputing facilities, and will profit enormously from the availability of quantum computers as special-purpose accelerators.
Quantum algorithms like the one presented here have the potential to solve many of the problems that plague today's simulation of correlated materials on classical computers, revolutionizing the field by opening new horizons for computational investigation of quantum materials.

A different class of algorithms that have been explored over the last few years are "variational quantum eigensolvers"~\cite{peruzzo2014,wecker2015variational}. These approaches, which are tailored for first-generation quantum computers that can only execute short circuits coherently, rely on Hamiltonians with only very few non-commuting terms to be practically feasible. As such, they are well-suited to the approximate simulation of simple, local model Hamiltonians such as the single-band Hubbard model. However, simulating complex materials as discussed here will be prohibitive both in the number of required qubits to represent an extended system and in the number of relevant non-commuting interaction terms, which severely affect the scaling of the variational algorithms.

During the completion of this work, a related approach where a quantum computer is used as impurity solver within the variational cluster approach~\cite{potthoff2003self,potthoff2003vca} appeared in Ref.~\onlinecite{dd2015}.
Our approach differs in several crucial ways. Most importantly, the embedding method used in our paper is the more broadly applicable and widely used DMFT method. Furthermore, we have described a zero-temperature approach in this manuscript, while the method of \cite{dd2015} operates at finite temperature. Since the circuit-based quantum computers for which both approaches
are designed perform unitary evolution of a quantum system, computing the required time-dependent correlation functions for thermal (mixed) states incurs significant overhead both in the number of
required qubits as well as the computation time. No attempt at estimating the scaling of the algorithm, or establishing a baseline for the gate counts, is made in Ref.~\onlinecite{dd2015}.

\section*{Acknowledgements}
The authors acknowledge useful discussions with Lei Wang and Ara Go.
Part of this research was performed at the Aspen Center for Physics supported by NSF grant \#PHY--1066293.
MT was supported by Microsoft Research, the European Research Council through ERC Advanced Grant SIMCOFE and the Swiss National Science Foundation through NCCR QSIT.

\appendix

\section{Introduction to Quantum circuits}
\label{app:circuits}

\begin{figure}
  \includegraphics{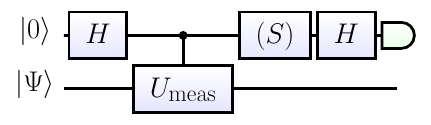}
  \caption{Incoherent measurement circuit for $U_\mathrm{meas}$. Here, $S=\sqrt{Z}$ is applied only if the imaginary part of the
  expectation value is desired. \label{fig:Umeas} }
\end{figure}

\begin{figure*}
  \centering
  (a) \\
  \includegraphics{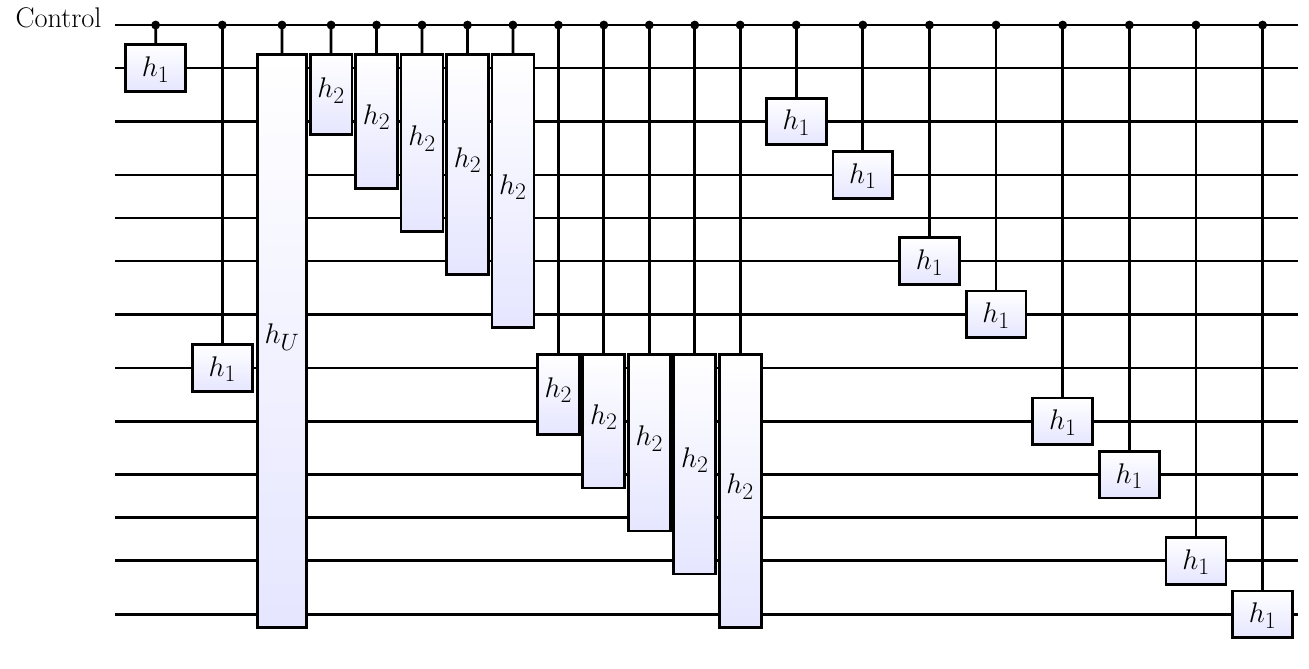} 
  
  \begin{tabular}{ccc}
  (b) &(c) &(d) \\
  \includegraphics{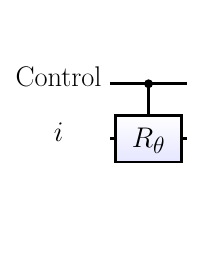} &\includegraphics{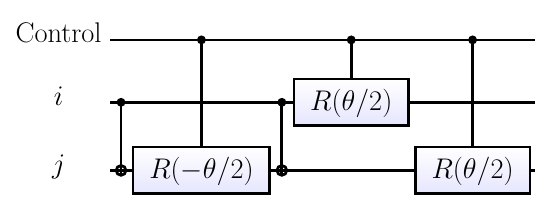} &\includegraphics{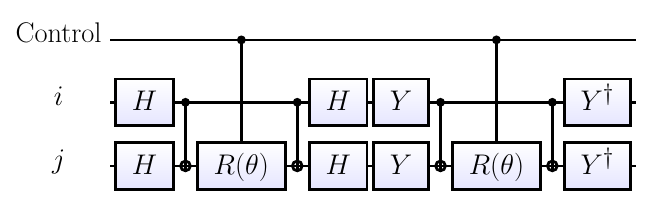}
  \end{tabular}
  
  \caption{
  (a): High-level overview of one step of the Trotterized time evolution.
  (b): Circuit $h_1 = \exp(-i T \varepsilon c_i^\dagger c_i)$.
  (c): Circuit $h_U = \exp(-i T U n_i n_j)$, where $\theta = TU/2$.
  (d): Circuit $h_2 = \exp(-i T t [c_i^\dagger c_j + c_j^\dagger c_i])$. Here we show the simplest case $j=i+1$, which does not require
  a Jordan-Wigner transformation.
  \label{fig:circuits}
  }
\end{figure*}

In this section, we show the quantum circuits necessary to implement the operations discussed in the main text. In 
these circuit diagrams, the horizontal lines indicate individual
qubits and boxes indicate quantum gates. For example, in the following circuit, the Hadamard gate $H=|\rightarrow\rangle\langle\uparrow| + |\leftarrow\rangle\langle\downarrow|$, which transforms between the $X$ and $Z$ basis, is applied to one qubit:
\begin{equation}
\includegraphics{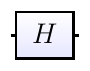}
\end{equation}
Other important ingredients are controlled gates, as shown in this circuit:
\begin{equation}
\includegraphics{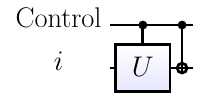}
\end{equation}
Here, we first apply a controlled unitary, and then a controlled-NOT (CNOT) gate. The operation ($U$ or NOT$=X$) is
applied to those components of the input state where the control qubit is in state $|1\rangle$, but not to those where the control
is in state $|0\rangle$.
Other gates we use are
\begin{align}
  R(\theta)&= \left( \begin{array}{cc} 1 &0 \\ 0 &e^{i \theta} \end{array} \right)
&Y&= \frac{1}{\sqrt{2}} \left( \begin{array}{cc} 1 &i \\ i &1 \end{array} \right)
\end{align}

Fig.~\ref{fig:Umeas} shows how to perform the measurement of a unitary $U_\mathrm{meas}$ by applying the unitary controlled on an ancilla qubit
that is in the state $1/\sqrt{2}(|0\rangle+|1\rangle)$. This will entangle the ancilla qubit with the qubits on which the unitary is applied
and make the expectation value accessible by measuring the ancilla.

Fig.~\ref{fig:circuits} shows how a single time step of the Trotter evolution is implemented as a quantum circuit.
In the present Hamiltonian, terms fall into three categories: (i) chemical potential terms of the form $h_1 = \varepsilon c_i^\dagger c_i$, (ii) hopping terms of the form $h_2 = t (c_i^\dagger c_j + c_j^\dagger c_i)$, and (iii) an interaction term $h_U = U n_i n_j$. Here, the subscript is a multi-index that contains both spin and orbital index. Fig.~\ref{fig:circuits} shows the way these different terms are implemented as quantum circuits. Here, for sake of completeness, we always include the control qubit that is necessary to perform subsequent steps of the algorithm.
In the case of the hopping circuit $h_{pq}$, in general a Jordan-Wigner transformation has to be performed to correctly account for the fermionic sign structure. For an overview of how to achieve this, we refer to Ref.~\onlinecite{hastings2014}.

\section{Quantum Algorithms}
\label{app:algorithms}

\subsection{Quantum simulation of time evolution}
The most basic building block of our simulation algorithm is the ability to simulate time dynamics of quantum systems on a quantum computer.
To this end, we first map the Hilbert space of the quantum system onto that of the quantum computer. The simplest method is to allocate one qubit per spin-orbital and work in a second-quantized occupation-number basis, i.e. use the state of the qubit to indicate whether the spin-orbital is occupied or empty. Next, we need to apply the unitary $\exp(-i H t)$ to this state. While many approaches to approximating this unitary on a quantum computer are
known~\cite{trotter1959,suzuki1976,childs2003exponential,anmer2011,childs2011simulating,anmer2012,berry2012black,childs2012hamiltonian}, we use here the simple approach of a Trotter-Suzuki decomposition~\cite{trotter1959,suzuki1976}. We decompose the Hamiltonian $H$ as a sum of non-commuting terms $H=\sum_i H_i$, where the $H_i$ include both one-body and two-body terms, and make the approximation
\begin{equation}
e^{-i H t} \approx \Bigl( e^{-i H_1 t/N} e^{-i H_2 t/N} \ldots\Bigr)^N,
\end{equation}
where $N$ is some integer. This approximation becomes exact as $N \rightarrow \infty$, and it may often be advantageous
to use higher-order decompositions~\cite{childs2012product}.
The simple quantum circuits that implement $\exp(-i H_i t/N)$ for the various terms $H_i$ are discussed in Appendix~\ref{app:circuits}. For the evolution under a time-dependent
Hamiltonian, as required for the adiabatic state preparation, we update the parameters of the Hamiltonian in each of the $N$
time steps.

\subsection{Quantum phase estimation}
Given the ability to apply $\exp(-i H t)$ to the state, we can measure the energy $\langle \psi |H| \psi\rangle$ of a given state $|\psi\rangle$
using an approach known as quantum phase estimation~\cite{kitaev1995,kitaev2002book}.
%This procedure avoids measuring the individual terms of the Hamiltonian
%separately, since such measurements do not commute among themselves and with $H$ and would thus destroy the state. In contrast,
%quantum phase estimation performs a measurement that is diagonal in the energy basis. It will also project a state close to the ground state
%onto the ground state with high probability. This is achieved with an accuracy $\epsilon \sim \mathcal{O}(1/T)$, where $T$ is the total computation time.
The basic idea of quantum phase estimation is to implement an interference experiment: an additional ancillary qubit (labelled PE) is used to control the application of $\exp(-i H t)$ so that if the PE-qubit is in the state $|1\rangle$, then $\exp(-i H t)$ is applied, while if it is $|0\rangle$, the identity operation is applied to the state. Effectively, this interferes two distinct trajectories with a phase difference $\exp(-i E t)$ where $E$ is the energy of the state. This phase difference rotates the angle of the qubit and by measuring this angle one can determine the phase difference. By taking large $t$, one can make the phase difference sensitive to small energy differences, allowing precise measurement of the energy by combining the information from measurements at several different times.

\subsection{Measurement of Green's functions} \label{app:gfct}
Within the computational model assumed here, we can perform measurements on individual qubits in a fixed basis.
In order to measure the real-time Green's function~\eqnref{eqn:rtgf}, we first relate its expectation values to those
of unitary operators~\cite{wecker2015}. We can then use a standard approach that allows the measurement of
expectation values of unitary operators. This approach, shown in detail in Appendix~\ref{app:circuits}, uses a controlled unitary operation
to entangle an ancilla qubit with the qubits on which the unitary should be measured, and thereby makes the expectation
value accessible through a measurement of the ancilla in the computational basis.

We define the unitary operators
\begin{equation} \label{eqn:Umeas}
U_\mathrm{meas}^{\alpha \beta}(t) = e^{i t H} q_\alpha e^{-i t H} q_\beta
\end{equation}
where
\begin{align}
q_1 &= c+c^\dagger &
q_2 &= i(c-c^\dagger).
\end{align}
When applied to an eigenstate $\ket{\Psi_n}$ with energy $E_n$, the unitary can be simplified using
$\bra{\Psi_n} e^{i t H} q_\alpha e^{-i t H} q_\beta \ket{\Psi_n} = e^{itE_n} \bra{\Psi_n} q_\alpha e^{-i t H} q_\beta \ket{\Psi_n}$.
In general, \eqnref{eqn:Umeas} contains terms of the form $c(t) c(0)$, $c^\dagger(t) c(0)$, etc. However, assuming
the absence of superconductivity, operators that do not conserve particle number will have vanishing expectation values in the
ground state and we obtain
\begin{align} \begin{split}
\eval{U_\mathrm{meas}^{11}} &= \eval{c(t) c^\dagger(0)} + \eval{c^\dagger(t) c(0)} \\
\eval{U_\mathrm{meas}^{12}} &= i \left( \eval{c^\dagger(t) c(0)} - \eval{c(t) c^\dagger(0)} \right).
\end{split} \end{align}
We can thus reconstruct the desired expectation values as
\begin{align} \begin{split}
G^p(t) &= \left( \eval{U_\mathrm{meas}^{11}} + i \eval{U_\mathrm{meas}^{12}} \right)/2 \\
G^h(t) &= \left( \eval{U_\mathrm{meas}^{11}} - i \eval{U_\mathrm{meas}^{12}} \right)/2.
\end{split} \end{align}
To get the real and imaginary part of both $G^p(t)$ and $G^h(t)$ for a given $t$ requires 4 measurements.
In the presence of superconductivity, where fermion number conservation is broken down to fermion parity conservation,
cross-terms such as $c(t)c(0)$ and $c^\dagger(t) c^\dagger(0)$ do not vanish
when evaluated on the ground state. Finding $G^p$ and $G^h$ thus requires measurement of the real and imaginary
part of all of the four operators $U^{\alpha \beta}_\mathrm{meas}(T)$, thus increasing the total
number of measurements at each time point from 4 to 8.

\section{Self-consistency and bath fitting}
\label{app:selfc}

The self-consistency loop which is used to determine the free parameters of the bath, $\epsilon_i$ and $V_{\alpha i}$,
is most conveniently and reliably executed in imaginary (Matsubara) frequencies, and
we must therefore extract the Green's function in imaginary time from our quantum impurity solver.
In the following, we will omit orbital indices for notational simplicity; in the general case, the Green's function
must be assumed to be a matrix.
We define the particle and hole contribution to the real-time Green's functions (cf Eqn.~\eqnref{eqn:rtgf}) as
\begin{align} \label{eqn:rtgfsupp} \begin{split}
G^p(t) =& \langle \Psi | c(t) c^\dagger(0) | \Psi \rangle \\
G^h(t) =& \langle \Psi | c^\dagger(t) c(0) | \Psi \rangle.
\end{split} \end{align}
The standard time-ordered Green's function $G(t) = -i \langle \Psi| \mathcal{T} c(t) c^\dagger(0) |\Psi \rangle$, where $\mathcal{T}$
is the time-ordering operator, can be recovered as % (using $\langle \Psi | c(t) c^\dagger(0) |\Psi\rangle = \langle \Psi | c(t+t') c^\dagger(t') |\Psi\rangle$) as
\begin{align} \begin{split}
G(t) = & -i \Theta(t) \langle \Psi | c(t) c^\dagger(0) \rangle \\ 
+ &i \Theta(-t)\langle \Psi | c^\dagger(-t) c(0) |\Psi \rangle \\
= -&i \Theta(t) G^p(t) + i \Theta(-t) G^h(-t)
\end{split} \end{align}
Performing a Fourier transform on the Green's function, $G(\omega) = \int_{-\infty}^\infty e^{i \omega t} G(t) dt$, 
we find using the above definitions:
\begin{align}
\begin{split}
G(\omega) = & -i \left[ \int_\epsilon^\infty dt\ e^{i (\omega+i \eta) t} G^p(t) \right.
	\\ &\left. + \int_\epsilon^\infty dt\ e^{-i(\omega-i \eta) t} G^h(t) \right],
\end{split}
\end{align}
where the lower bound $\epsilon$ in the time integrals has been introduced for later convenience, and can be taken
to be on the order of the floating point precision when numerically performing the integrals. $\eta$ is a numerical broadening
factor that should be taken small compared to the relevant physical energy scales of the system for extracting the spectral function,
but can be taken to be $\eta=0$ if only the imaginary-frequency Green's function is desired.
Viewed as a function of complex frequencies $z=\omega+i \omega_n$, the many-body Green's function $G(z)$ is analytic in
the lower and upper complex half-plane, with non-analyticities only along the real axis $\operatorname{Im} z = 0$, and asymptotically
behaves as $G(z) \rightarrow 1/|z|$ for $|z| \rightarrow \infty$.
Following standard definitions, the spectral function is given by
\begin{align}
A(\omega) = & -2 \im G(\omega) = i \left( G(\omega) - \bar{G}(\omega) \right)
\end{align}
In this definition of the spectral function, positive frequencies encode the particle contribution, and negative frequencies encode the hole contributions.
For equilibrium systems, $A(\omega) \geq 0$.

Given the Green's function on the real frequency axis, or the spectral function, we can rely on the analyticity properties mentioned
above and use a Hilbert transformation of the spectral function to obtain the Green's function in imaginary frequencies:
\begin{equation}
G(i \omega_n) = \int_{-\infty}^\infty \frac{d \omega}{2 \pi} \frac{A(\omega)}{i\omega_n - \omega}.
\end{equation}
Using the integrals (for $t > 0$, $\omega_n > 0$)
\begin{align} \begin{split}
\int_{-\infty}^\infty \frac{d \omega}{2\pi} \frac{e^{i \omega t}}{i \omega_n-\omega} &= -i e^{-t \omega_n} \\
\int_{-\infty}^\infty \frac{d \omega}{2\pi} \frac{e^{-i \omega t}}{i \omega_n-\omega} &= 0,
\end{split} \end{align}
one can for example compute the particle contribution to be
\begin{gather*}
G^p(i \omega_n) = \\
\int_\epsilon^\infty \frac{dt\ e^{-\eta t} }{2 \pi} \int_{-\infty}^{\infty} \frac{d\omega }{i \omega_n - \omega} \left[ e^{i \omega t} G^p(t) - e^{-i \omega t} \bar{G}^p(t) \right] \\
= -i \int_0^\infty dt\ e^{-t(\eta+\omega_n)} G^p(t).
\end{gather*}
Here, $\bar{G}$ denotes complex conjugation. The computation for the hole contribution proceeds analogously, to yield the final expression (for $\omega_n > 0$)
\begin{equation}
G(i \omega_n) = -i \int_{\epsilon}^\infty dt\ e^{-t(\eta+\omega_n)} \left[G^p(t) + \bar{G}^h(t) \right]. \label{eqn:suppGiw}
\end{equation}
We turn this into a discrete sum over a set of times $t_i$ at which the integrand is evaluated. For best convergence of the
integral, we use an improved integrator scheme such as Simpson's rule. We choose the times as
\begin{align}
t_i &= \epsilon \exp\left(\log(T/\epsilon) \frac{i}{N-1} \right) & i &= 0,\ldots,N-1.
\end{align}

We now describe the DMFT self-consistency condition, which relates the free parameters of the impurity model to the parameters
of the original Hubbard model. Here, we follow closely the notation of Section VI.A.1.d of Ref.~\onlinecite{georges1996}.
The self-consistency condition can be stated as
\begin{equation} \label{eqn:Gnew}
G(i \omega_n) = \int_{-\infty}^\infty d\epsilon\ \frac{D(\epsilon)}{i\omega_n+\mu-\Sigma(i\omega_n)-\epsilon}
\end{equation}
where $\mu$ is the chemical potential, $\Sigma$ denotes the self-energy,
\begin{equation} \label{eqn:Sigma}
\Sigma(i\omega_n) = G_0^{-1}(i \omega_n) - G^{-1}(i\omega_n),
\end{equation}
$G_0$ is the non-interacting impurity Green's function corresponding to a set of bath parameters, and $G$ is the impurity Green's function
measured from the solution of the interacting impurity problem. In general, the properties of the bath are encapsulated in the hybridization
function $\Delta(i \omega_n) = \Sigma(i \omega_n) + G^{-1}(i \omega_n) - i \omega_n - \mu$. For the specific case of a discrete bath used in this paper,
the hybridization function is related to the bath parameters by $\Delta(i \omega_n) = \sum \frac{V_i^2}{i \omega_n - \epsilon_i}$, and the non-interacting
Green's function is then given by~\cite{krauth1994}
\begin{equation} \label{eqn:G0discr}
\left( G_0^\mathrm{discr}(i \omega_n) \right)^{-1} = i\omega_n + \mu + \sum_i \frac{V_i^2}{i \omega_n - \epsilon_i}.
\end{equation}

In most practical cases, the self-consistency must be achieved iteratively by means of executing the self-consistency loop.
Given as input the non-interacting Green's function $G_0$ (which in the case of a discrete bath as discussed here is parametrized
by the bath parameters $\epsilon_i$, $V_i$ through~\eqnref{eqn:G0discr}) as well as the interacting impurity Green's function $G$ obtained from the solution
of the impurity problem for that set of bath parameters, one first computes the self-energy $\Sigma$ using~\eqnref{eqn:Sigma} and then evaluates
\begin{equation}
\widetilde{G_0}^{-1} = \left( \int_{-\infty}^\infty \frac{d\epsilon\ D(\epsilon)}{i\omega_n+\mu-\Sigma-\epsilon} \right)^{-1} + \Sigma
\end{equation}
to obtain a new estimate for the non-interacting Green's function.

Having obtained the new non-interacting Green's function in imaginary frequencies, we must obtain the bath
parameters $\epsilon_i$, $V_i$ (for a single spin-degenerate impurity) such that the discrete version of the non-interacting Green's function of Eqn.~\eqnref{eqn:G0discr}
best approximates the desired Green's function.
To this end, we optimize the cost function
\begin{equation} \label{eqn:costf}
\sum_n \left| \widetilde{G_0}^{-1}(i \omega_n) - \left( G_0^\mathrm{discr}(i \omega_n) \right)^{-1} \right|^2
\end{equation}
using a non-linear optimization scheme in the parameters $V_i$, $\epsilon_i$. For more details
on how to reliably perform optimization of the bath parameters, see Ref.~\onlinecite{go2015}.

In the case of the Bethe lattice~\cite{bethe1935statistical} in the limit of infinite coordination number, the density of states follows the semicircular
form $D(\epsilon)=\frac{1}{2\pi} \sqrt{4-\epsilon^2}$, $-2 \leq \epsilon \leq 2$, and 0 otherwise (here we set the hopping integral
to $t=1$)~\footnote{See, e.g., Ref.~\onlinecite{georges1996}}. In this particular case, the integral of~\eqnref{eqn:Gnew}
can be evaluated to find (where $z = i \omega_n + \mu - \Sigma(i \omega_n)$)
\begin{equation}
G(i \omega_n) = \frac{1}{2} \left( z- \sqrt{z^2-4} \right).
\end{equation}
Using~\eqnref{eqn:Sigma}, this can be simplified to yield the concise self-consistency condition for the Bethe lattice~\cite{krauth1994}:
\begin{equation} \label{eqn:selfcBethe}
G_0^{-1}(i \omega_n) = i\omega_n + \mu - G(i \omega_n).
\end{equation}
In this case, a new non-interacting Green's function $\widetilde{G_0}^{-1}$ can be obtained by simply evaluating the right-hand
side of~\eqnref{eqn:selfcBethe} using the numerically obtained impurity Green's function $G$. This is then used as input for the
same bath fitting procedure of~\eqnref{eqn:costf}.

\section{Gate-count estimates}
\label{app:gatec}

\subsection{Incoherent approach}

The most naive possible workflow to measure an expectation value for a fixed time $t$ is as follows:
\begin{enumerate}
\item Prepare the ground state $\ket{\Psi_0}$ by adiabatic evolution from an easily prepared initial state, for example free fermions
or the atomic limit.
\item Measure the expectation value of the unitary operator $U_\mathrm{meas}^{\alpha \beta}$. This is achieved using the circuit illustrated
in Fig.~\ref{fig:Umeas}, which projects into a final state $\ket{\Psi_1}$.
\end{enumerate}
Since the most costly step of the above procedure is the preparation of $\ket{\Psi_0}$, it would be very helpful to
reduce the number of preparations. This seems feasible since the measurement only measures a single bit of information,
and the final state will likely still have significant overlap with the ground state. To exploit this, we can apply
a projective measurement that, if successful, projects the state back into the ground state. This can be achieved
by performing QPE on the unitary $U=e^{-itH}$ to measure the energy. In the successful
case, where the final measurement in the QPE yields the known ground state energy, the state has been successfully
projected into the ground state and can be used as input to a new measurement; however, if the measurement yields
an eigenstate of different energy, which is orthogonal to the ground state, the state would have to be re-prepared.
To avoid this, the measurement of the ancilla qubits in the final step of QPE can be replaced
by the measurement of a single qubit which encodes whether the system is in the ground state or not.
This yields the algorithm sketched in Fig.~2 of the main manuscript. The probability of returning to the ground state, which
dictates how often the state must be re-prepared from scratch, depends on the numerical details of the model and must
therefore be estimated on a case-by-case basis.

Numerical simulations of the example discussed in the main text show that following the above scheme, we
need to prepare the ground state from scratch $1.7 \cdot 10^5$ times, and
have to perform QPE for a total of $6.8 \cdot 10^6$ times.

\subsection{Coherent approach}

The coherent measurement procedure was first described in
Ref.~\onlinecite{wecker2015}. For a detailed description of this approach,
we refer to this reference. For the purpose of this paper, it suffices to note that this approach gains a quadratic speedup
in the accuracy, i.e. the number of measurements required at each time step are reduced quadratically.

\bibliography{qc4mat}

\end{document}